\documentclass[prl, twocolumn]{revtex4-1}

\usepackage{natbib}
\usepackage{graphicx}
\usepackage{epstopdf}
\usepackage{amsmath, amssymb}
\usepackage{color}
\usepackage{xspace}

\def \e { \mbox{$\mathrm{e}$} }

\newcommand\eg{\textit{e.g.}\xspace}
\newcommand\ie{\textit{i.e.}\xspace}

\renewcommand\Re{\mbox{$\mathrm{Re}$}}

\newcommand{\vkt}[1]{\tilde{#1}}
\newcommand{\ensbl}[1]{\left\langle #1 \right\rangle}
\newcommand\vkf{\vkt{f}}
\newcommand \vkS{\vkt{S}}
\def \vth {\mbox{$v_{\scriptsize{\mathrm{T}}}$}}
\def \omegaNL {\mbox{$\omega_{\mbox{\tiny{NL}}}$}}
\def \gammaEff {\mbox{$\gamma_{\mbox{\scriptsize{eff}}}$}}

\begin{document}

\title{Landau Damping in a Turbulent Setting}

\author{G. G. Plunk}
\email{gplunk@ipp.mpg.de}
\affiliation{Max-Planck-Institut f\"{u}r Plasmaphysik, EURATOM-Assoziation, Wendelsteinstr. 1, 17491 Greifswald, Germany}

\begin{abstract}
To address the problem of Landau damping in kinetic turbulence, the forcing of the linearized Vlasov equation by a stationary random source is considered.  It is found that the time-asymptotic density response is dominated by resonant particle interactions that are synchronized with the source.  The energy consumption of this response is calculated, implying an effective damping rate, which is the main result of this paper.  Evaluating several cases, it is found that the effective damping rate can differ from the Landau damping rate in magnitude and also, remarkably, in sign.  A limit is demonstrated in which the density and current become phase-locked, which causes the effective damping to be negligible; this potentially resolves an energy paradox that arises in the application of critical balance to a kinetic turbulence cascade.
\end{abstract}

\maketitle

\section{Introduction}

Though linear Landau damping \cite{landau-r, *landau-e, vk, case} is completely understood for an isolated Fourier mode, it is not understood precisely how this phenomenon manifests in settings where many modes or degrees of freedom are interacting chaotically by some nonlinear coupling, \ie in a turbulent setting.  It is not clear if Landau's long-time solution is useful at all in this context.  Indeed, this solution corresponds to the most weakly damped root among a hierarchy of roots, and only emerges after a transient period.  For a Fourier mode subject to sustained forcing, there is no justification for assuming that it will ever be found in this asymptotic state, and thus no reason to suppose that the Landau damping rate will determine how quickly energy is removed from the system.

In this paper we investigate how linear Vlasov dynamics respond to a random source.  We interpret this source as a representation of arbitrarily strong turbulent interactions.  This scenario is in contrast to wave turbulence of the ``fluid'' type, \ie where the state of the system is described by a small number of nonlinearly interacting fields that support linear wave solutions.  The random forcing of the corresponding linearized fluid system is typically not a very interesting problem to consider, since the response can be anticipated as the resonant response of the wave solutions, and the low-dimensional intuition of driven oscillators applies.  The problem becomes more interesting when one considers exotic linear dynamics such as that of non-normal operators \cite{farrell}, where the eigenspectrum by itself is an incomplete description of the linear dynamics \cite{trefethen-book}.  In the case of a collisionless plasma, the wave solutions are replaced by an infinite hierarchy of modes, and the intuition from low-dimensional forced linear systems no longer applies.  Thus, the problem of forced linearized Vlasov equation is inherently interesting, and, apparently, largely unexplored.  Considering that the linear solutions given by the Landau roots are all exponentially damped, a na\"{i}ve guess would be that damping in the nonlinear state could be determined by an appropriately weighted average of the Landau rates.  For the forced Vlasov equation, we find that this guess completely misses the mark.


The calculation made here is a logical step in line with the statistical formulation of turbulence, which is not concerned with individual solutions of the nonlinear dynamics, but rather seeks to understand generic or universal properties of the ensemble.  In light of the subtle linear dynamics of the Vlasov equation, it should not be surprising that the conclusions here are physically nontrivial.  It is important to emphasize that it is a crucial simplification to assume that the dynamics are essentially random; but this assumption clearly also limits the applicability of the solution.  In particular, the general problem of nonlinear Landau damping, as addressed in the celebrated work of \citet{mouhot-villani}, is clearly beyond the scope of this approach.

The results of the calculation are summarized as follows.  We find that under sustained forcing by a stationary random source, Landau damping is supplanted by another process, namely the resonant response of the particles to the source.  This interaction causes the distribution function to tend toward coherent steady-state solutions that consume energy at a rate different from the Landau damping rate.  Depending on the frequency spectrum of the source and other parameters, the effective damping rate (defined in terms of this energy consumption) can be much smaller, comparable, or much larger than the Landau damping rate; in some cases the energy can even flow inversely, corresponding to a {\em negative} effective damping rate.  This occurs in the absence of linearly unstable eigenmodes, and thus may be thought of as an example of ``generalized instability'' \cite{farrell} in the context of a continuum of stable modes.  This is a context in which generalized instability has been expected by some authors to play an insignificant role in the generation of turbulence, and thus the findings here constitute a possible counterexample to such expectations.

Our findings shed new light on the phenomenon of ``critical balance'' \cite{goldreich} in kinetic turbulence.  Originally formulated for magnetohydrodynamic (MHD) turbulence, critical balance can be extended to a kinetic context \cite{schekochihin-apj}.  The basic hypothesis of critical balance is that the turbulence cascade will proceed anisotropically through scales where the linear mode frequency matches an effective nonlinear turnover frequency.  For kinetic turbulence, however, the presence of Landau damping seems at odds with the existence of a cascade at all.  That is, basic considerations of energy balance imply that a critically balanced local cascade, subject to scale-by-scale damping (at the rate predicted by linear Landau damping), must suffer a loss of energy flux as the cascade proceeds to smaller scales \cite{howes2008}.  Thus, one might expect the energy to dissipate rather than cascade.  In this paper, we resolve this paradox by demonstrating that critically balanced fluctuations can be effectively undamped, even when the Landau rate is strong.  This may explain recent numerical simulations that demonstrate the existence of critically balanced cascades \cite{howes2011, barnes-cb-itg, tenbarge2012}.

The solution presented in this work may also help point the way to advanced Landau-fluid models \cite{hammett-perkins}, which have been the subject of sustained interest for their promise to significantly simplify the analysis of kinetic systems.  The central question in this context is how to model the interactions between a hierarchy of fluid moments and how to truncate this hierarchy in a manner which retains the important physics of the fully kinetic system.  The forced solutions of this paper constitute exact analytic solutions that can be used to test the capabilities of Landau-fluid systems.

\section{Formulation of the Problem}

Landau damping, as it was first calculated \cite{landau-r, *landau-e}, describes how plasma perturbations damp according to a continuous description (the Vlasov equation), in the absence of any model of collisional dissipation.  In this form, it is a basic example of how irreversible behavior can occur in a fundamentally reversible linear system.  However, discretized models including small but non-zero dissipation have been shown to yield the same solution \cite{ng-bhattacharjee-skiff-prl, swanson2003}, but with the added conclusion that the damping must be attributed to the specific physical mechanism of collisions.  By contrast, damping in the (collisionless) Vlasov description relies on the continuum to provide a limitless repository for structure in velocity space, with successively smaller and smaller scales forming as damping proceeds.  It is a comforting fact that these formulations ultimately yield the same answer.  In this paper, we will work within the collisionless continuum description, as it is elegant and succinct, and avoids non-universal features associated with finite collisionality.

A popular explanation of Landau damping appeals to the image of plasma receiving energy from the electric field by ``surfing'' on waves -- with a bit more plasma traveling slower than the wave than that traveling faster, a net energy transferred from the wave to the plasma, thus damping the wave.  However, though energy balance clearly must be satisfied, the general solution attributed to Case \cite{case} and Van Kampen \cite{vk, vk-book} reveals that linear Landau damping is actually due to a systematic smearing or ``phase mixing'' of the distribution function that is formally equivalent to the free evolution of a population of uncoupled harmonic oscillators with a distribution of frequencies.  (An alternate but equivalent statement is that the stable linear system can be transformed to action-angle variables \cite{morrison-pfirsch, morrison-aa}.)  The analogy between oscillator populations and plasmas has fueled fruitful interaction between the fields \cite{strogatz2000}.  For nonlinearly coupled oscillators, one finds both ordered and disordered states, with disordered states exhibiting Landau damping and ordered states corresponding to phase-locking or ``synchronization'' of the oscillators.  It is interesting to note that we also find both damping and synchronization in the present study, establishing an intriguing qualitative connection.

As an example of the kind of system we would like to understand, consider the electrostatic slab drift-kinetic equation.  This equation describes plasma dynamics in the presence of a strong uniform magnetic guide field ${\bf B} = \hat{\bf z} B$ in a five-dimensional phase space ($x$, $y$, $z$, $v_{\parallel}$, $v_{\perp}$), where $v_{\parallel} = \hat{\bf z}\cdot{\bf v}$ and $v_{\perp} = |\hat{\bf z}\times{\bf v}|$.  Fourier-transforming in position space and defining $k_{\parallel} = \hat{\bf z}\cdot{\bf k}$ we have

\begin{equation}
\frac{\partial f}{\partial t} + i k_{\parallel} v_{\parallel} f + i(k_{\parallel}v_{\parallel} - \omega_*)\varphi F_0 = \displaystyle{\sum_{{\bf k}^{\prime}} } \epsilon({\bf k}, {\bf k}^{\prime}) \varphi({\bf k}^{\prime}) f({\bf k} - {\bf k}^{\prime}),\label{drift-kinetic-eqn}
\end{equation}

\noindent where $\varphi$ is the normalized electrostatic potential satisfying $n_0 \varphi = 2\pi \alpha\int dv_{\parallel}v_{\perp}dv_{\perp}\; f$, $\epsilon({\bf k}, {\bf k}^{\prime}) = \epsilon_0(\hat{\bf z}\times{\bf k}^{\prime})\cdot{\bf k}$ is the nonlinear coupling coefficient, and $\omega_*({\bf k}, v_{\perp}, v_{\parallel})$ is a linear frequency, which depends on scale lengths of the background density $n_0$ and temperature $T_0$.  The distribution $f$ represents small plasma fluctuations about a large background $F_0$, satisfying $2\pi\int dv_{\parallel}dv_{\perp} F_0 = n_0$.  Noting that none of the terms proportional to the distribution function have any dependence on $v_{\perp}$ in their coefficients, we can integrate over this variable.  We then take the nonlinear term as a given function, denoted $S({\bf k}, v_{\parallel}, t)$.  Finally, as there is no explicit reference to $k_x$ or $k_y$, we suppress these variables and henceforth only refer to a single scalar wavenumber $k = k_{\parallel}$ and velocity variable $v = v_{\parallel}$ and substitute the notation $2\pi\int v_{\perp}dv_{\perp} F_0 \rightarrow F_0(v)$.

\section{Formal solution}

The subject of what follows is a very simple equation for a single Fourier component of the distribution of particles, $f(k, v, t)$.  The wavenumber $k$ is now just a parameter and the problem takes the form of a continuum of oscillators of natural frequency $kv$ that interact via an order parameter, the density $n$.  That is, we are left with the one dimensional forced Vlasov equation

\begin{equation}
\frac{\partial f}{\partial t} + i k v f + i k c_s^2 n(t) G(v) = S(v, t),\label{plasma-eqn}
\end{equation}

\noindent where $c_s$ is a characteristic wave propagation speed in the plasma.  The function $S(v, t)$ is the source, which we take to represent interaction with a large number of other Fourier components that compose a bath of turbulent fluctuations.  The third term on the left hand side of Eqn.~\ref{drift-kinetic-eqn} has been rewritten in terms of a general function $G(v)$; we note that the uniform-background ($\omega_* = 0$) single-species case considered by Landau, it is defined $G(v) =  -\partial_v F_0/n_0$.  The fluctuation density $n(t)$ is defined

\begin{equation}
n(t) = \int_{-\infty}^{\infty} f (v, t) dv.\label{n-eqn}
\end{equation}
 
\noindent Eqn.~\ref{plasma-eqn} is a reduced kinetic description in the sense that there there is only one velocity dimension and a single scalar wavenumber (dependence on the full wavevector ${\bf k}$ is concealed in constants and can also be accounted for in the definition of $G(v)$).

To solve this equation, we perform an invertible transformation \cite{morrison-pfirsch, morrison-aa} that is implied by the solution of Van Kampen \cite{vk-book}.  (Note that the solution may be sought by other well-established methods, \eg via the Vlasov linear response function \cite{krommes2002}, or by Laplace transform.)  Thus, Eqn.~\ref{plasma-eqn} is equivalent to

\begin{equation}
\frac{\partial \vkf}{\partial t} + i ku \vkf = \vkS(u, t),\label{plasma-eqn-trans}
\end{equation}

\noindent where the transformed distribution is defined

\begin{equation}
\vkf(u) = \frac{f_{+}(u)}{D_{+}(u)} + \frac{f_{-}(u)}{D_{-}(u)},
\end{equation}

\noindent and we define $D_{\pm}(u) = 1 \pm 2\pi i c_s^2 G_{\pm}(u)$.  The positive- and negative-frequency parts of an arbitrary function $g(v)$ are defined in terms of the Fourier transform by

\begin{equation}
g_{\pm}(u) = \pm \int_0^{\pm \infty} d\nu \e^{i \nu u} \int_{-\infty}^{\infty} dv\frac{\e^{-i \nu v}}{2 \pi} g(v).
\end{equation}

\noindent Note that $g(u) = g_{+}(u) + g_{-}(u)$ and $D_{+} = D_{-}^*$ with $*$ denoting the complex conjugate.  Also, note that the absence of the interaction term in Eqn.~\ref{plasma-eqn-trans} now makes our problem that of forced non-interacting harmonic oscillators.  This equation may be solved in the frame rotating at the oscillator frequency $-ku$ by directly integrating with respect to time.  We find

\begin{equation}
\vkf(u, t) = \e^{- i ku t} \vkf_0(u) + \e^{- i ku t} \int_0^{t} dt^{\prime} \e^{i ku t^{\prime}} \vkS(u, t^{\prime}),\label{vk-soln}
\end{equation}

\noindent where $\vkf_0(u) = \vkf(u, 0)$.  The inversion of our transformation is achieved by the formula

\begin{equation}
f(v, t) = \int_{-\infty}^{\infty} du \vkf(u, t) f^{u}(v),\label{vkt-inverse}
\end{equation}

\noindent where $f^{u}(v)$ is an eigenmode (\eg a Case-Van Kampen mode) of the unforced system (Eqn.~\ref{plasma-eqn} with $S = 0$), defined $f^{u}(v) = \lambda(u)\delta(u - v) + c_s^2 P [G(v)/(u - v)]$ and $P$ denotes the principal value with respect to the point $u = v$.  The eigenmodes are normalized such that $\lambda(u) = 1 - c_s^2 P\int_{-\infty}^{\infty} dv G(v)/(u - v)$.  Plugging our solution into this formula and using Eqn.~\ref{n-eqn} we find the following expression for the density:

\begin{equation}
n(t) = \int_{-\infty}^{\infty} du \;\e^{- i ku t} \left[ \vkf_0(u) + \int_0^{t} dt^{\prime} \e^{i k u t^{\prime}} \vkS(u, t^{\prime})\right].\label{phi-full-soln}
\end{equation}

\noindent  The first term is easily recognized as the solution due to Van Kampen of the damping of an initially smooth distribution in velocity space.  This term tends to zero at large times $t \rightarrow \infty$ since the factor $\exp(-i kut)$ becomes increasingly oscillatory, \ie by the Riemann-Lebesgue lemma.  We are more interested in the time-asymptotic behavior and so we will focus on the second term.  If we consider the harmonically driven case, \ie $S(v, t) = \exp(-i \Omega t) \hat{S}(v)$, this term yields  

\begin{equation}
n_{\Omega}(t) = \e^{-i\Omega t}\int_{-\infty}^{\infty} du \left[\frac{1 - \e^{-i (k u - \Omega)t}}{i(ku - \Omega)}\right] \hat{\tilde{S}}(u).\label{harmonic-forced-soln-1}
\end{equation}

\noindent Note that for finite time the integrand is non-singular at $ku = \Omega$.  However, this quantity must be carefully evaluated as $t \rightarrow \infty$.  Defining $x = ku - \Omega$, the quantity in brackets can be written as

\begin{equation}
\left[\frac{1 - \e^{-ixt}}{ix}\right] = \pi \eta_{t}(x) - i \chi_{t}(x),
\end{equation}

\noindent where $\eta_{t}(x) = \sin(xt)/(\pi x)$ and $\chi_{t}(x) = (1 - \cos(xt))/x$.  We can then use the identities $\lim_{t \rightarrow \infty}\;  \eta_{t}(x) = \delta(x)$ and $\lim_{t \rightarrow \infty}\;  \chi_{t}(x) = P(\frac{1}{x})$ and $(i/\pi)P\int dx^{\prime} f(x^{\prime})/(x - x^{\prime}) = f_{+}(x) - f_{-}(x)$ to evaluate the time-asymptotic response of the density.  Thus for $t \rightarrow \infty$ the quantity $n_{\Omega}$ of Eqn.~\ref{harmonic-forced-soln-1} is evaluated as

\begin{equation}
\lim_{t \rightarrow \infty} n_{\Omega}(t) = \e^{- i\Omega t}\left(\frac{2\pi}{k}  \frac{\hat{S}_{+}(\Omega/k)}{D_{+}(\Omega/k)}\right),\label{harmonic-forced-soln-2}
\end{equation}

\noindent where we have assumed $k > 0$ for simplicity.  This solution is a purely oscillatory mode induced by resonant forcing of particles with $u = \Omega/k$.  The secular growth of the distribution function at this velocity point dominates over all other contributions at large times.  For a general forcing function $S(v, t) = \int_{-\infty}^{\infty} d\Omega \exp(-i \Omega t) \hat{S}(v, \Omega)$, we can simply integrate this response to obtain

\begin{equation}
\lim_{t \rightarrow \infty} n(t) = \frac{2\pi }{k} \int d\Omega \; \e^{- i \Omega t} \frac{\hat{S}_{+}(\Omega/k, \Omega)}{D_{+}(\Omega/k)} \label{t-asymp-soln}
\end{equation}

\noindent Note that for a stationary random source, this solution is undamped because the spectral components of $S$ are uncorrelated, \ie $\hat{S}$ is not a smooth function of $\Omega$.  Introducing the ensemble average $\ensbl{.}$, stationarity implies $\ensbl{\hat{S}_{+}(u, \Omega)\hat{S}_{+}^{*}(u^{\prime}, \Omega^{\prime})} = \delta(\Omega - \Omega^{\prime}) \Lambda(u, u^{\prime}, \Omega)$.  From Eqn.~\ref{t-asymp-soln} we find the ensemble response

\begin{equation}
\ensbl{|n|^2} = \frac{4\pi^2}{k^2}\int_{-\infty}^{\infty} d\Omega \; \frac{\Lambda(\Omega/k, \Omega/k, \Omega)}{|D_{+}(\Omega/k)|^2}.\label{n2-response}
\end{equation}

\noindent This is the ensemble response of the density.  Note the use of the time-asymptotic solution to compute the ensemble response.  This is valid if the time-scales present in the source are much smaller than the time domain of the system, so that the system spends most of its time in the asymptotic state.  

To grasp the physical meaning of Eqns.~\ref{t-asymp-soln} and \ref{n2-response} for an actual turbulent system, it may be helpful to consider the following {\it Gedankenexperiment}.  Imagine two boxes of plasma, plasma A and plasma B.  Plasma A constitutes a steady turbulent bath of fluctuations and Plasma B is initially uniform and quiescent.  Now imagine that we have a perfect measurement device capable of exactly resolving the features of the turbulent bath of Plasma A, and also a perfect source capable of driving the plasma B in an arbitrary fashion.  With this experimental setup, we measure the complete instantaneous Fourier spectrum of Plasma A and, choosing a specific Fourier mode, exactly reconstruct the signal $S(k, v, t)$ that is driving that mode.  Then, we use this signal to set the source as an input to the initially quiescent plasma.  After a short period of time, \ie a time comparable to the inverse of the characteristic frequency of $S(k, v, t)$, plasma B will exhibit precisely one Fourier mode at finite amplitude.  By measuring this mode and comparing it with the same mode in plasma A, we should find that the density moment of both will be given by Eqn.\ref{t-asymp-soln}, with possible corrections due to low frequency contributions to $S(k, v, t)$; other moments such as the current and temperature fluctuations should also agree with expressions analogous to Eqn.\ref{t-asymp-soln}.  Thus, for all practical purposes, the mode in Plasma B should be a ``clone'' of the mode in plasma A.  This correspondence depends on the fact that the influence of the initial condition will be lost to conventional Landau damping.  The accuracy of this reproduction also depends on the assumption that the frequency spectrum at fixed ${\bf k}$ is peaked to some degree about a characteristic frequency.  Fortunately, there is evidence that plasma turbulence does exhibit characteristic frequencies (see for instance \cite{mazzucato-nstx, gorler-jenko-pop}), even in a strongly nonlinear state.

We can now compute the average energy consumption rate of our solution.  By energy, we refer in this case to the quantity $|n|^2/2$, which only differs from the physical (electrostatic) energy by a constant.  Energy balance is found by multiplying Eqn.~\ref{plasma-eqn} by $n^*$, integrating over velocity, ensemble-averaging and taking the real part.  By stationarity we have $\partial_t \ensbl{|n|^2}/2 = 0$ and what remains is

\begin{equation}
\Re[ i k \ensbl{j n^*}] = \Re[\ensbl{n^* \int_{-\infty}^{\infty} dv S(v)}],\label{energy-balance}
\end{equation}

\noindent where we have defined the current $j = \int_{-\infty}^{\infty} v dv f$.  This equation expresses the average balance of energy input by the source and consumption by the linear phase-mixing term (\ie wave-particle interaction).  We define the energy input $\varepsilon = \Re[\ensbl{n^* \int_{-\infty}^{\infty} dv S(v)}]$, which we can evaluate using Eqn.~\ref{t-asymp-soln}.

\begin{equation}
\varepsilon = \frac{2\pi}{k}\int d\Omega\int dv \; \Re[ \frac{\Pi(\Omega/k, v, \Omega)}{D_{+}(\Omega/k)} ],\label{energy-consumption}
\end{equation}

\noindent where we define $\Pi$ via $\ensbl{\hat{S}_{+}(u, \Omega) \hat{S}^*(v, \Omega^{\prime})} = \delta(\Omega - \Omega^{\prime}) \Pi(u, v, \Omega)$.  Finally, let us define the effective damping rate implied by Eqns.~\ref{n2-response} and \ref{energy-consumption}.

\begin{equation}
\gammaEff = \frac{\varepsilon}{\ensbl{|n|^2}}.\label{effective-damping}
\end{equation}

\section{Examples}

The Landau damping rate $\gamma_L$ is determined by the velocity dependence of the background distribution function $F_0(v)$ and the ratio of the plasma velocity $c_s$ to the velocity of typical particles, \eg $\vth$.  The response given by Eqn.~\ref{n2-response} retains this dependence via the function $D_{+}$.  In fact the ``dispersion relation'' for the Landau roots is obtained from the zeros of the analytic continuation of $D_{+}$ into the lower half plane.  Our solution is also determined by the statistics of the turbulence, via the function $\hat{S}$.  Let us now examine some simple cases to show how $\gammaEff$ can differ from $\gamma_L$.

Equation \ref{effective-damping} is an exact result, only assuming stationarity, but depends on unknown statistics of the source.  Let us now take a very simple source, and then consider how the the result applies to a more general class of sources.  The goal is to determine generic properties of the solution, starting from specific examples that are relatively easy to follow.  

The spectrum of an appropriate turbulent bath should, for fixed ${\bf k}$, be peaked about a characteristic frequency $\Omega \sim \omegaNL$, as discussed above.  Thus, for simplicity let us consider a source with just one frequency, $S = \exp(-i \omegaNL t) f_M S_0$, where $f_M = \exp(-v^2/\vth^2)/(\sqrt{\pi}\vth)$ gives Maxwellian velocity dependence and $S_0$ is a randomly phased complex amplitude.  We substitute $\Lambda(\Omega/k, \Omega/k, \Omega) = \delta(\Omega - \omegaNL) |f_{M+}(\omegaNL/k)|^2 |S_0|^2$ and $\Pi(\Omega/k, v, \Omega) = \delta(\Omega - \omegaNL) f_{M+}(\omegaNL/k) f_M(v) |S_0|^2 $ and find

\begin{equation}
\gammaEff = \frac{k}{2\pi}\Re[\frac{D_{+}(\omegaNL/k)}{f_{M+}(\omegaNL/k)}],\label{effective-damping-2}
\end{equation}

\noindent were we note that $f_{M+}(\omegaNL/k) = Z(\zeta)/(2i\pi\vth)$, where $Z$ is the plasma dispersion function and $\zeta = \omegaNL/(k\vth)$.

\begin{figure}
\includegraphics[width=0.95\columnwidth]{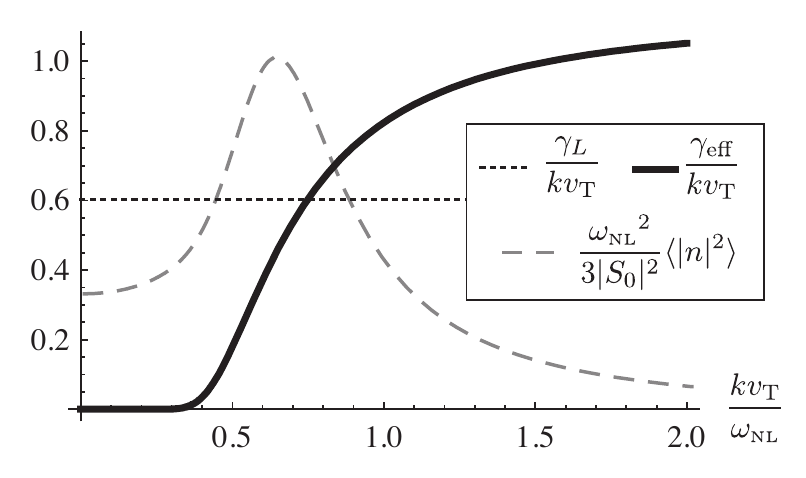}
\caption{Comparison of Landau damping rate with effective damping rate for the case of uniform background.  The Landau rate $\gamma_L$ is thin-dashed, the effective rate $\gammaEff$ is solid and $\alpha = 2 c_s^2/\vth^2 = 1$.  Also plotted in thick-dashed (gray) is the density response, normalized to fit the figure.  The Landau root is strongly damped with $\omega \approx (1.45 - 0.60 i)k\vth$.  This mode corresponds to an ion acoustic wave with an ion-electron temperature ratio of $1$.}
\label{gamma-compare-fig}
\end{figure}

As an example, let us first consider $G(v) = 2v f_M/\vth^2$.  This is the classic case considered by Landau, corresponding to a single species plasma with a spatially uniform background.  We evaluate $D_+(\omegaNL/k) = 1 + \alpha[1 + \zeta Z(\zeta)]$ where $\alpha = 2 c_s^2/\vth^2$.  Then the Landau damping rate is computed from the zeros of $D_+(\omega/k)$ in the lower half-plane.  We plot the effective damping rate compared with the Landau rate in Fig.~\ref{gamma-compare-fig}.  Imagining the plasma as a single effective oscillator with natural frequency $k\vth$, we can interpret the plotted quantity $\gammaEff/(k\vth)$ as the inverse of the quality factor of this oscillator.  We see that for high $k\vth/\omegaNL$ the quality factor tends to a constant of about 0.9, whereas for low $k\vth/\omegaNL$ the quality factor becomes infinite; this is because the density and current become phase-locked in this limit.

Fig.~\ref{gamma-compare-fig} shows that effective damping of the density response can be quite significant, even exceeding the Landau rate, but becomes negligible for $k\vth/\omegaNL < 0.5$.  Thus, if a cascade occurs that is not significantly damped, it seems reasonable to conclude that critical balance, now stated as the condition $k\vth/\omegaNL \lesssim 1$, must be satisfied.  Indeed, the numerical simulation of tokamak turbulence by \citet{barnes-cb-itg} (see Fig.~3) exhibits a spectrum of fluctuation energy that is dominated by $k\vth$ below the effective nonlinear frequency.

As a second example, let us now take $G(v) = 2f_M/\vth[x - \kappa\{1 + \eta(x^2 - 1/2)\}]$, where $x = v/\vth$, and $\kappa$ and $\eta$ are parameters.  This corresponds to the slab ion-temperature-gradient (ITG) mode, given the following definitions.  We take $\eta = L_n/L_T$ (where $L_n$ and $L_T$ are gradient scale lengths of the background density and temperature respectively) and $\kappa = k_y\rho/(k L_n)$ (where $\rho = \vth/(2\Omega_c)$, $\Omega_c$ is the ion cyclotron frequency and $k_y$ is the wavenumber perpendicular to the directions of the magnetic field and background gradients); Finite Larmor radius effects are neglected for simplicity.  The homogeneous system (Eqn.~\ref{plasma-eqn} with $S = 0$) is stable for $0 < \eta < 1+ \sqrt{2(1+\alpha)\kappa^{-2}\alpha^{-2} + 1}$ (see \eg \cite{goldston-rutherford}).  We find $D_+(\omegaNL/k) = 1 + \alpha[1 + \zeta Z(\zeta) + \kappa\{ (\eta/2 - 1)Z(\zeta) - \eta\zeta(\zeta Z(\zeta) + 1)\}]$.

\begin{figure}
\includegraphics[width=0.95\columnwidth]{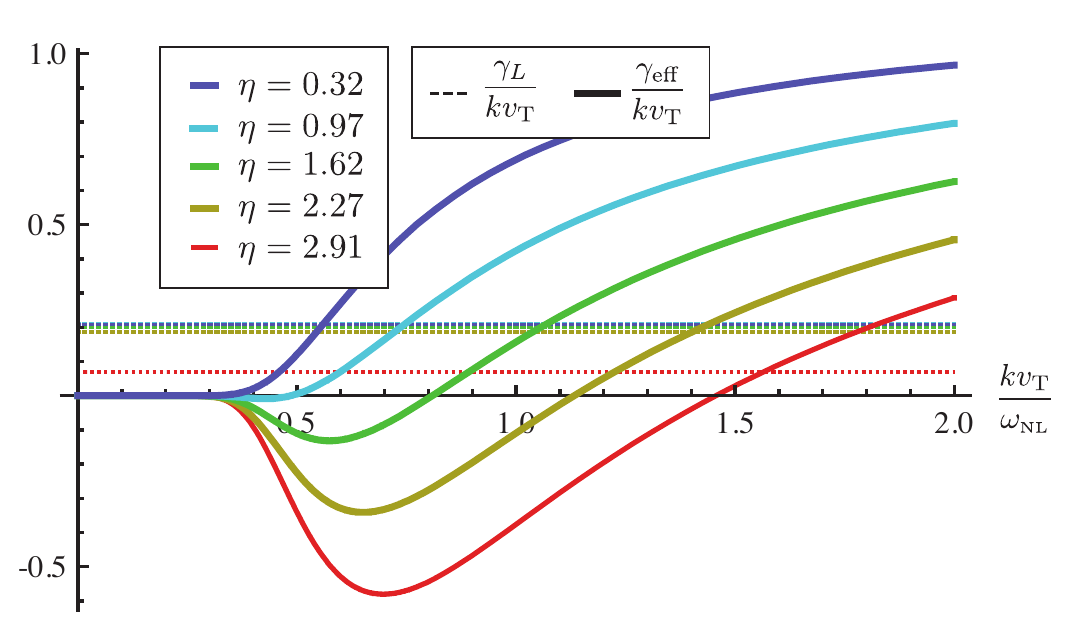}
\caption{Comparison of Landau damping rate with effective damping rate for several cases of the ITG system.  Parameters are $\kappa = 1$ and $\alpha = 2c_s^2/\vth^2 = 1$; all cases are sub-marginal, $\eta < 1+ \sqrt{5}$.  Landau rates $\gamma_L$ are dashed lines, effective rates $\gammaEff$ are solid lines.  The plots are in order, with $\eta$ increasing from top to bottom.}
\label{gamma-compare-fig-ITG}
\end{figure}

We plot a few cases in Fig.~\ref{gamma-compare-fig-ITG}.  Here we see a similar behavior as the homogeneous case but find that the effective damping can actually take on negative values as $\eta$ approaches the marginal point of the ITG instability.  The difference of this behavior from the behavior of the Landau solution is striking, and also points to the intriguing possibility that the ITG mode could sustain turbulence in the absence of linear instability.

\section{Final remarks}

Although we have assumed the special case of Maxwellian forcing to arrive at Eqn.~\ref{effective-damping-2}, the absence of damping at low $k\vth/\omegaNL$ is actually more general than it might appear.  In fact, we may just as well make the substitution $f_M(v) \rightarrow F(v)$ in our forcing, where $F(v)$ is a purely real function that falls off rapidly in $v$ -- \eg a Maxwellian function multiplied by a polynomial in $v$.  Then $F_+$ would appear in place of $f_{M+}$ in Eqn.~\ref{effective-damping-2}.  Using a well known identity (see \eg \cite{vk-book}), we may write $F_+ = (F + i F_*)/2$, where $F_*(v) = \pi^{-1}P\int_{-\infty}^{\infty} dv^{\prime}F(v^{\prime})/(v-v^{\prime})$ is the Hilbert transform of $F$.  Because $F$ is real, the real and imaginary parts of $F_+$ are $F/2$ and $F_*/2$, respectively.  Consequently, at large argument, the real part of $F_+$ goes rapidly to zero while the imaginary part decays algebraically; the latter can be verified by a multipole expansion of the Hilbert transform.  Thus, the asymptotic behavior of $F_+$ is qualitatively the same as that of $f_{M+}$ and so the effective damping under this forcing will rapidly go to zero at sufficiently small $k\vth/\omegaNL$.  

If, on the other hand, we consider a forcing function whose complex phase depends on $v$, finite damping can occur at small $k\vth/\omegaNL$.  Physically, positive damping in this case would result if a relative phase (of the correct sign) is favored between the current and density moments of the source.  However, there is no asymmetry in the fundamental equations to cause a preferred phase difference -- that is, the linear eigenmodes of the autonomous system ($S = 0$) have no relative phase (this reflects the time-reversal symmetry of the equation, which is also obeyed by the nonlinear system, Eqn.~\ref{drift-kinetic-eqn}).  For a turbulent cascade, such a phase difference may indeed be represented in the range where energy is injected (indeed, unstable ITG modes exhibit a phase difference between the density and current moments), but there is no reason, {\em a priori}, to expect this phase difference in the so-called inertial range, where it is commonly observed that symmetries of the dynamical equations are satisfied ``statistically'' \cite{frisch}.  Furthermore, note that if we assume that the bath of fluctuations, from which the source $S$ is composed (see Eqn.~\ref{drift-kinetic-eqn}), has no statistically preferred phase difference, then our solution for $f$ will likewise present no phase difference, and thus a system having this property is self-consistent.



The author acknowledges fruitful discussions with J. Krommes, P. Helander, J. Parker, A. Schekochihin and D. Schwab.  Thanks also to the Wolfgang Pauli Institute for hosting a series of workshops on gyrokinetics, which greatly stimulated this work.

\bibliography{critical-synchrony}

\end{document}